\documentclass[11pt]{article}
\usepackage{blois,epsfig}
\usepackage{multirow}

\def\Journal#1#2#3#4{{#1} {\bf #2}, #3 (#4)}

\def\PRL{\em Phys. Rev. Lett.}

\def\be{\begin{equation}}
\def\ee{\end{equation}}
\def\bea{\begin{eqnarray}}
\def\eea{\end{eqnarray}}

\newcommand{\met}       {\mbox{\ensuremath{\slash\kern-.7emE_{T}}}}
\newcommand{\dzero}     {D\O}
\begin{document}
\vspace*{4cm}
\title{Single top quark production and Vtb at the Tevatron}

\author{ {\large} Reinhard Schwienhorst \\
On behalf of the D0 and CDF collaborations
}

\address{Department of Physics and Astronomy, Michigan State University, 
East Lansing, MI 48826, USA}

\maketitle\abstracts{
Single top quark production via the electroweak interaction was observed by the 
D0 and CDF collaborations at the Tevatron proton-antiproton collider at Fermilab.
Multivariate analysis techniques are employed to extract the small single 
top quark signal. The combined Tevatron cross section is $2.76^{+0.58}_{-0.47}$~pb.
This corresponds to a lower limit on the CKM matrix element $|V_{tb}|$ of 0.77. 
Also reported are measurements of the $t$-channel cross section, the top quark
polarization in single top quark events, and limits on gluon-quark flavor-changing 
neutral currents and $W'$~boson production.
}

\section{Introduction}
The observation of single top quark production via the electroweak interaction was first reported
in 2009 by the \dzero~\cite{Abazov:2009ii} and CDF~\cite{Aaltonen:2009jj} collaborations. 
Both experiments reported observation at the five standard deviation level and measured
the CKM matrix element $V_{tb}$~\cite{Cabibbo:1963yz,Kobayashi:1973fv}. 
Their results were also combined~\cite{Group:2009qk}.
Here we present the two observation analyses, their combination, and several
searches for new physics in the single top quark final state.

\section{Cross section measurement}
 Single 
top quark production proceeds via the $t$-channel exchange of a virtual $W$~boson between
a light quark line and a heavy quark line or the $s$-channel production and decay of a virtual 
$W$~boson. The single top quark observation analysis from the \dzero~collaboration was done in the lepton
(electron or muon) +jets final state using 2.3~fb$^{-1}$ of data. The CDF collaboration utilized
both the lepton+jets final state with 3.2~fb$^{-1}$ of data and the missing transverse energy ($\met$)
+jets final state with 2.1~fb$^{-1}$ of data~\cite{Aaltonen:2010fs}. \dzero~later added the 
$\tau$ lepton +jets final state~\cite{Abazov:2009nu}.

The single top quark final state contains a lepton and a neutrino from the $W$~boson decay,
a $b$~quark from the top quark decay and an additional light quark ($t$-channel) or $b$~quark
($s$-channel). The event selection for all analysis channels requires large $\met$ and two to 
four jets (\dzero) or two to three jets (CDF), at least one of which is $b$-tagged. The analysis 
channels differ in the lepton requirement from the $W$~boson decay. The channels with the largest 
sensitivity require exactly one electron or exactly one muon. CDF has an additional channel requiring no 
reconstructed lepton, and \dzero~has an additional channel requiring a reconstructed $\tau$~lepton. In total,
about 13,000 events are selected by the two collaborations with an expected SM signal of about 550 events.

\begin{figure}
\psfig{figure=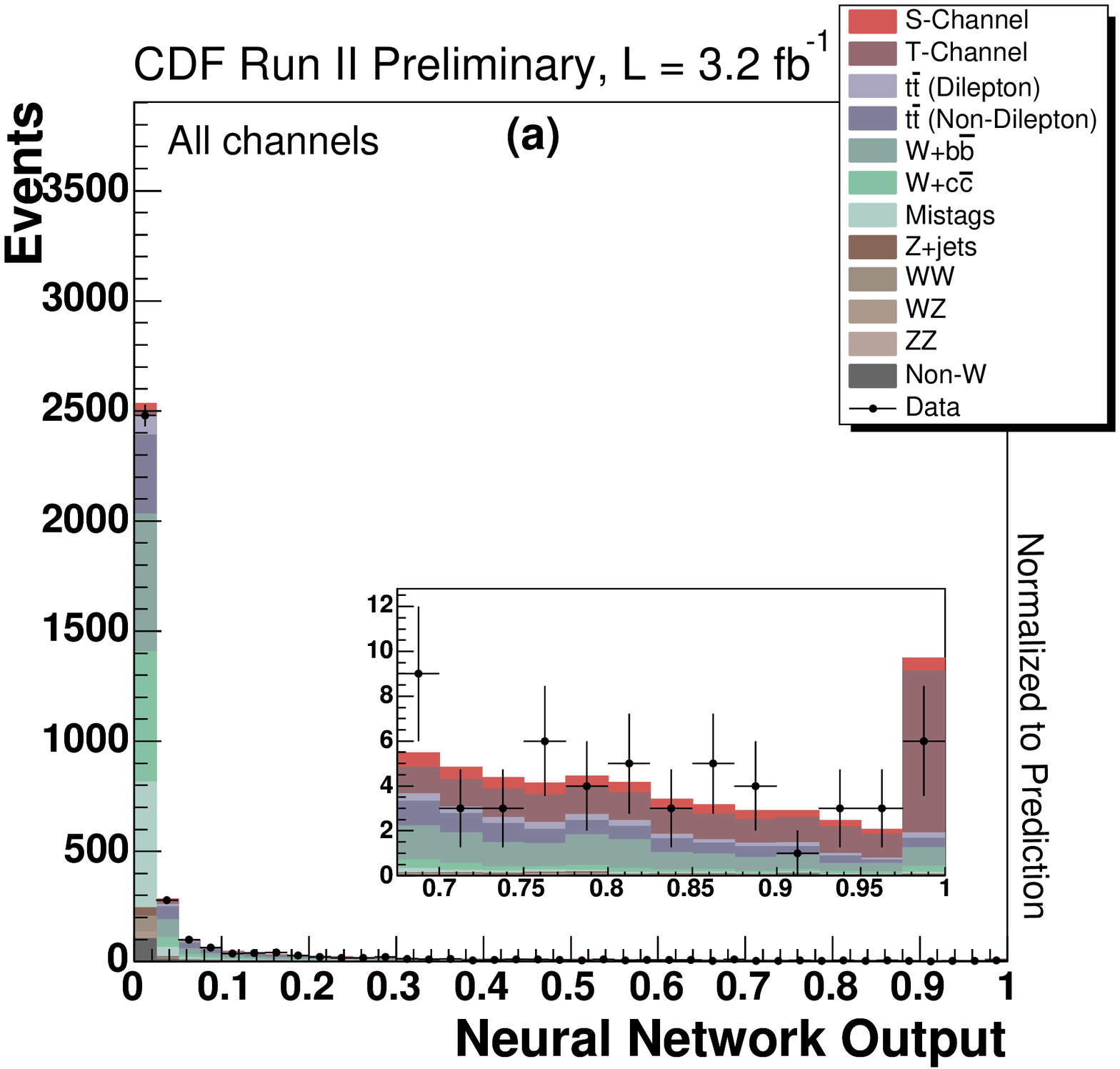,width=0.38\textwidth}
\psfig{figure=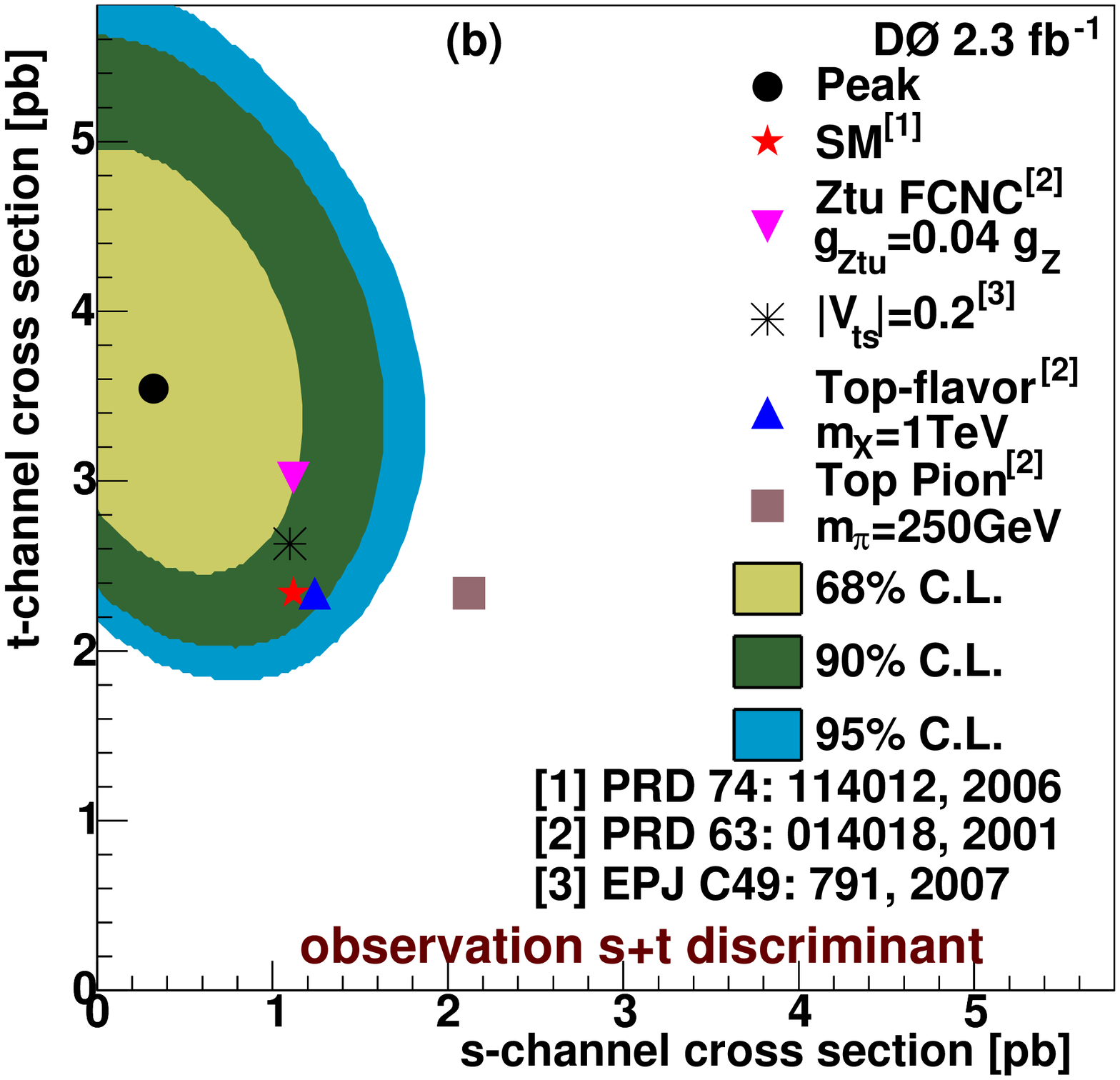,width=0.33\textwidth}
\psfig{figure=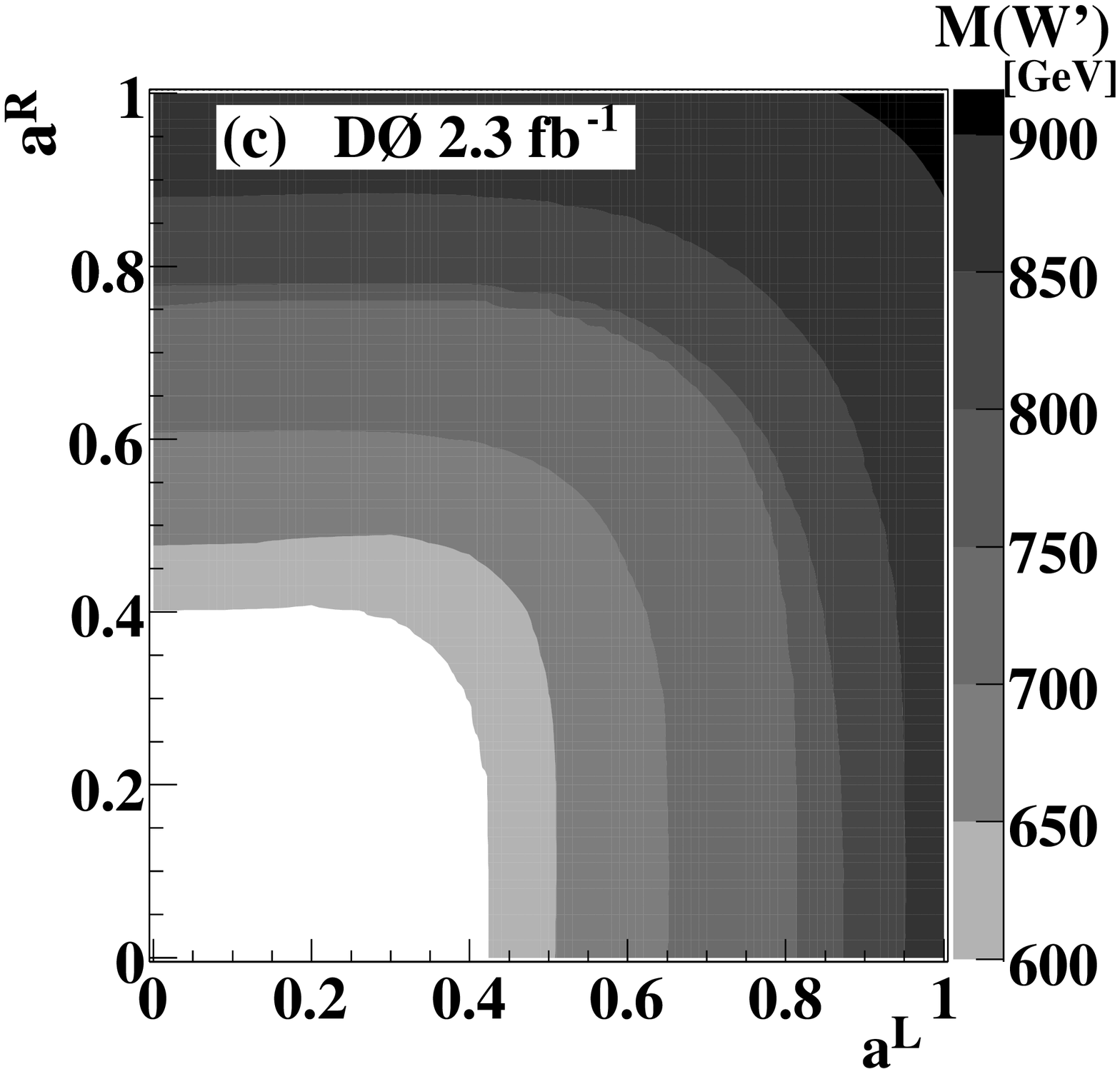,width=0.27\textwidth} 
\caption{(a) The combination neural network output of the CDF single top quark observation, (b) the posterior
probability density of the \dzero~single top quark observation as a function of the $s$-channel and $t$-channel
cross sections, and (b) limit on the mass of the $W'$~boson as a function of the left-handed and right-handed
$W'$~boson coupling.
\label{fig:xs}}
\end{figure}

The expected single top quark signal is small compared to the statistical uncertainty and the
uncertainty on the background prediction. Thus, multivariate analysis methods are employed to extract
the single top quark signal. \dzero~and CDF both use several different multivariate methods which are then 
combined into one filter. 
The output of the combined single top filter for the CDF experiment is shown in Figure~\ref{fig:xs}(a). The
single top quark signal is clearly visible above the backgrounds in the high-discriminant region. 
~

Both \dzero~and CDF observed single top quark production at the five standard deviation level. The measurement 
of the single top quark production cross section from \dzero~is 3.94$\pm$0.88~pb, and the measurement
from CDF is $2.3^{+0.6}_{-0.5}$~pb. Figure~\ref{fig:xs}(b) shows the \dzero~Bayesian posterior probability density 
distribution as a function of both the $s$-channel
and the $t$-channel cross sections. Also shown are
the SM expectation~\cite{stxs-harris} and several new physics models that affect single top quark 
production~\cite{Tait:2000sh,Alwall:2007}.
The combined Tevatron cross section is $2.76^{+0.58}_{-0.47}$~pb~\cite{Group:2009qk}. 

The single top quark production cross section measurement also provides a measurement of the CKM matrix 
element $V_{tb}|$, without requiring any assumtions about the number of generations or the unitarity of the 
CKM matrix~\cite{Cabibbo:1963yz,Kobayashi:1973fv}. The combined Tevatron measurement of $|V_{tb}|$ is 
$0.88\pm0.07$, with a lower 95\% confidence level limit of 0.77, using a theory cross section of 
$\sigma^{theory}_{s+t}=3.46|V_{tb}|^2$ by Kidonakis~\cite{stxs-kidonakis}.

The \dzero~collaboration isolated the $t$-channel mode of single top quark production by training multivariate 
filters for this signal alone. The result is first evidence for single top quark production in the 
$t$-channel at the 4.8 standard deviation level. The corresponding $t$-channel production cross section, 
making no assumptions about the $s$-channel cross section, is $\sigma_t=3.1\pm0.9$~pb~\cite{Abazov:2009pa}. 

The top quark decay width and lifetime is extracted from a combination of this measurement and a 
measurement of the top quark decay branching ratio~\cite{Abazov:2008yn}. The measured width is 
$\Gamma_t=1.99^{+0.69}_{-0.55}~\mathrm{GeV}$,
which corresponds to a top quark lifetime of $\tau_t =(3.3^{+1.3}_{-0.9})\times10^{-25}~{\rm s}$.

\section{Searches for new physics}
The single top quark final state is sensitive to many new physics signatures. The $s$-channel final state 
could receive contributions from a new heavy $W'$~boson, whereas the $t$-channel could receive 
contributions from flavor-changing neutral current (FCNC) interactions. 
All single top quark signatures are sensitive to modifications of the
top quark coupling to the $W$~boson and the $b$~quark.

\subsection{Search for $W'$ boson production}
Models of new physics that introduce additional symmetries predict the existence of additional heavy 
charged bosons, generally called $W'$. 
The \dzero~collaboration performed a search for $W'$~boson production in 2.3~fb$^{-1}$ of data, 
using the same event selection and background modeling
as the single top quark observation analysis. This is the first $W'$~boson search in the single top quark
final state that explores the full parameter space of both SM-like left-handed couplings and right-handed 
couplings and a mixture of both for the $W'$~boson. The resulting limits on the $W'$~boson mass as a 
function of the two couplings are shown in Figure~\ref{fig:xs}(c). For SM-like left-handed $W'$~couplings, 
the limit on the $W'$~boson mass is $M(W^\prime)>863$~GeV. For purely right-handed couplings it depends on
whether a right-handed neutrino ($\nu_R$) exists and is lighter than the $W'$~boson, the limit is 
$M(W^\prime)>885$~GeV for $M(W^\prime) < m(\nu_R)$ and 
$M(W^\prime)>890$~GeV for $M(W^\prime) > m(\nu_R)$. If both left-handed and right-handed couplings
are present, the limit is $M(W^\prime)>916$~GeV.
~
\begin{figure}
\begin{tabular}{cc}
 & \multirow{2}{*}{
\psfig{figure=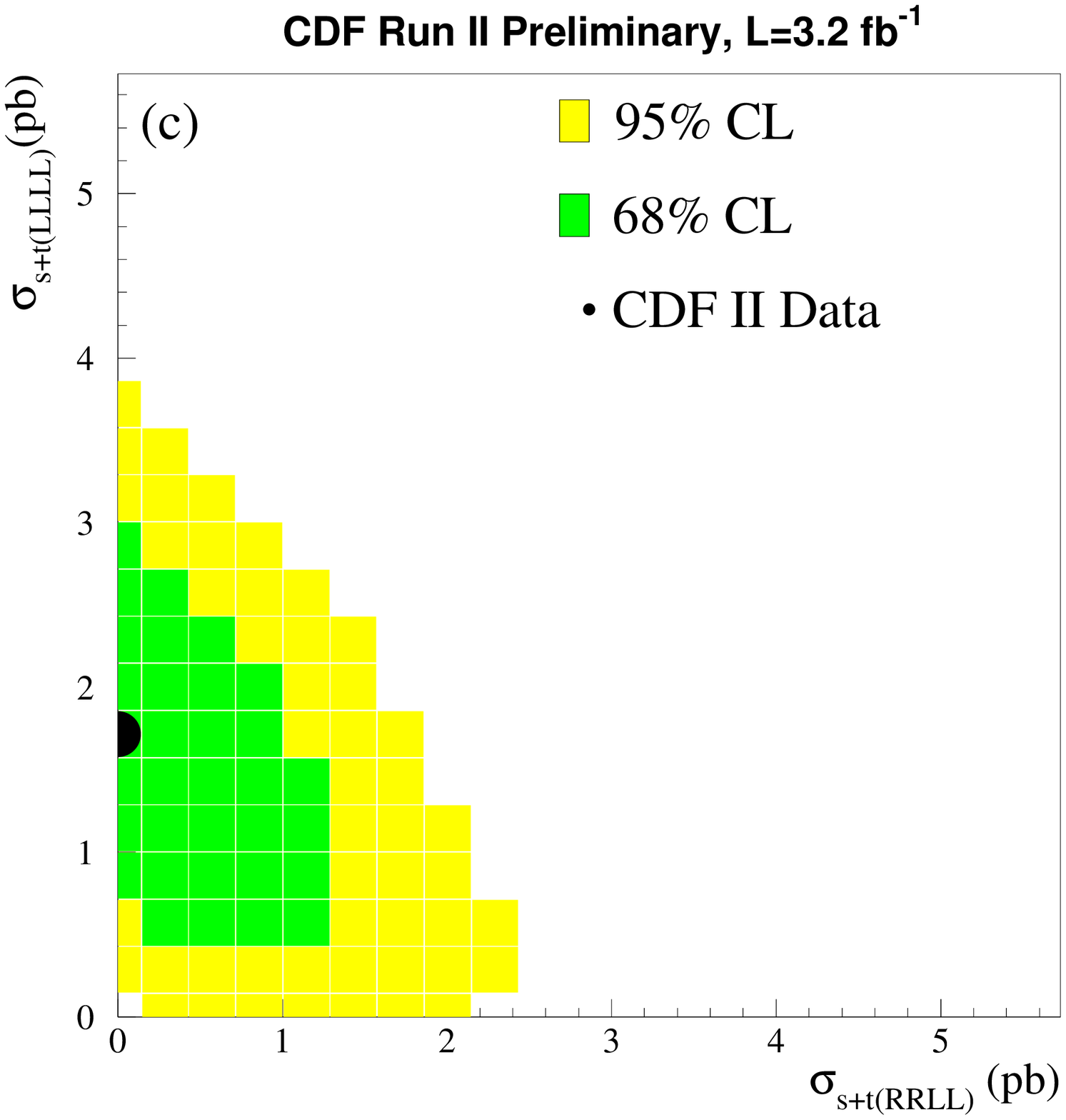,height=3.1in} } \\
\psfig{figure=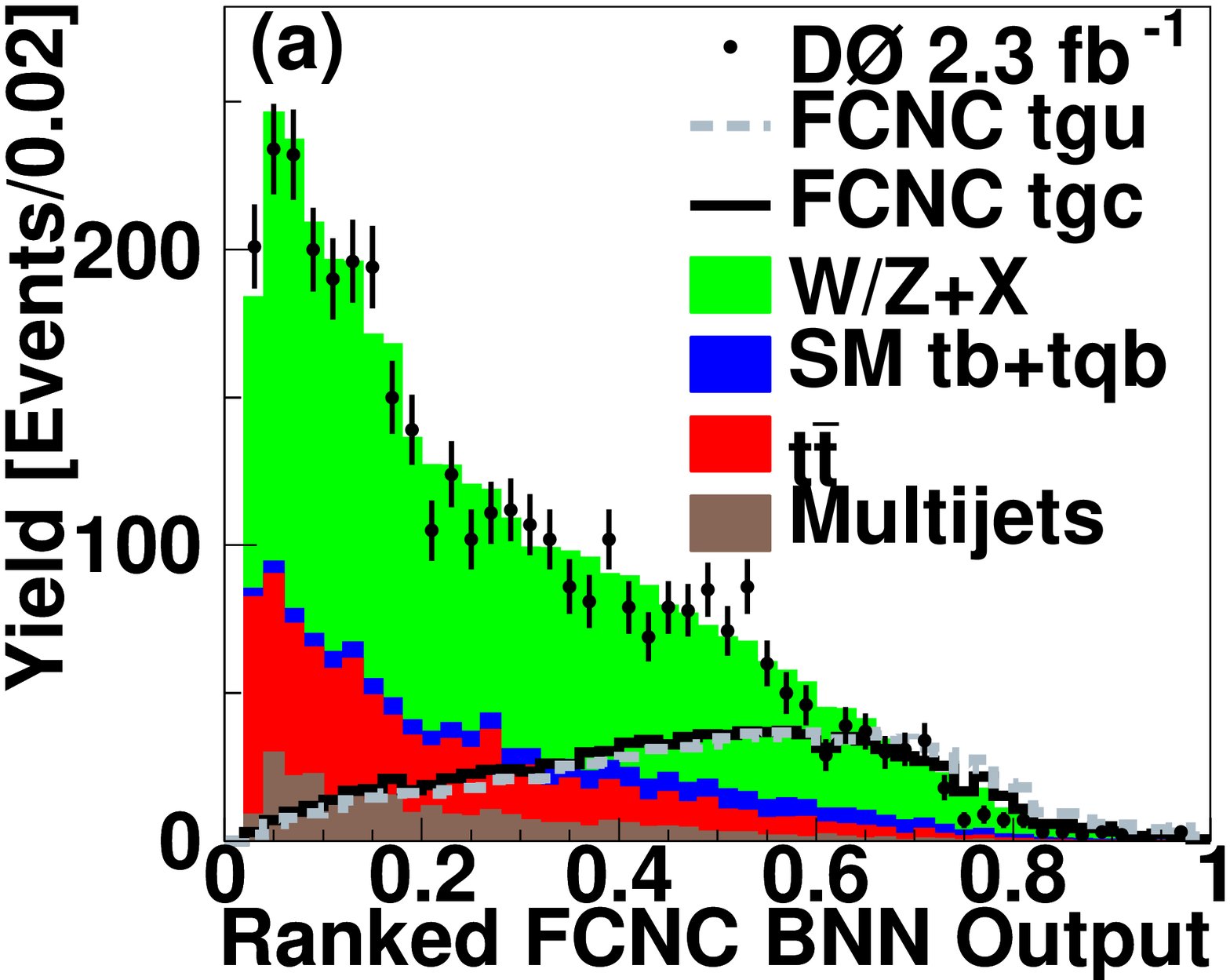,width=0.28\textwidth} & \\
\psfig{figure=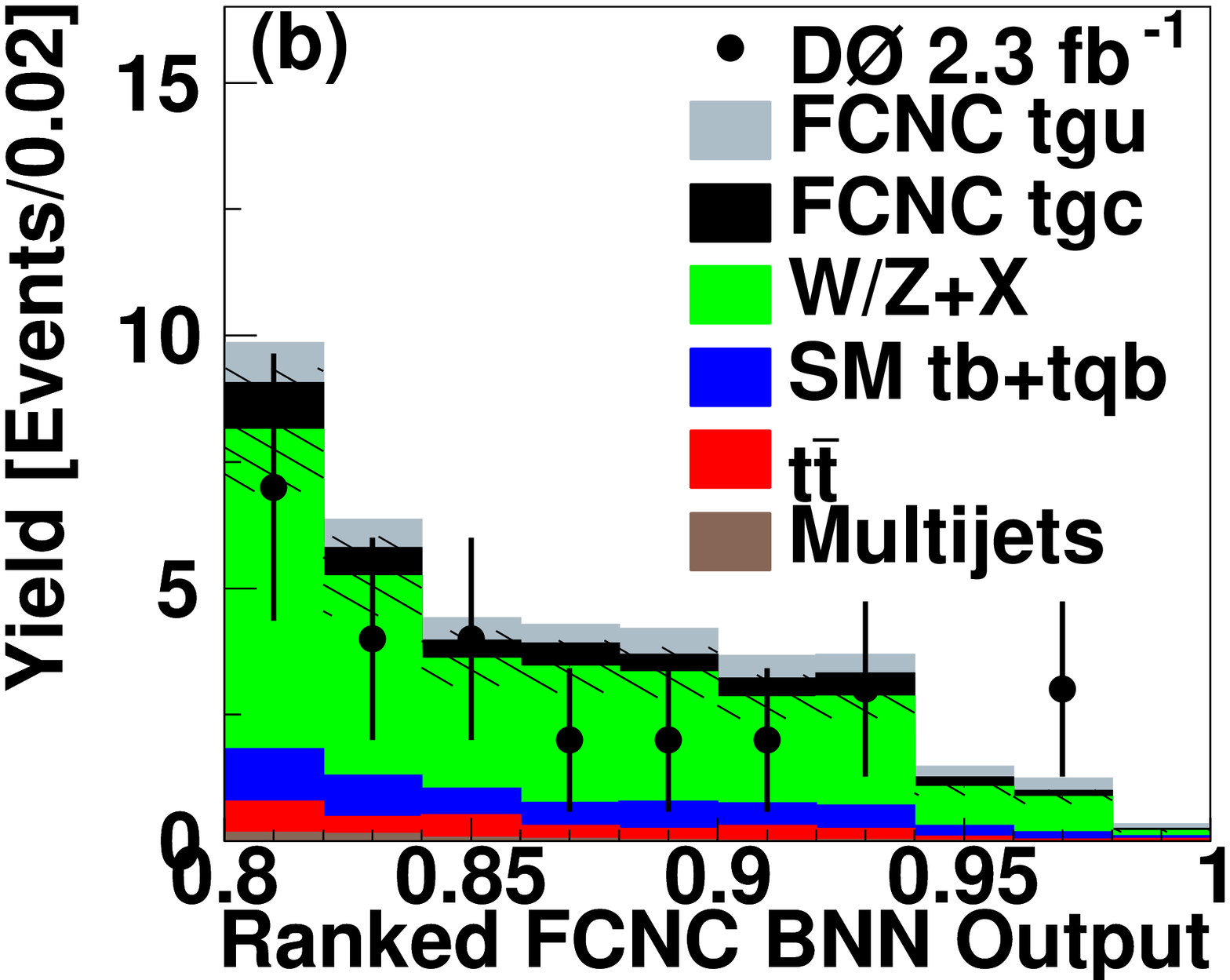,width=0.28\textwidth} & \\
\end{tabular}
\caption{ (a) and (b) comparison of the background model to data for the FCNC 
discriminant summed over all analysis channels, (a) for the whole discriminant range and (b) only 
the high discriminant region, where the hatched region gives the uncertainty on the background sum. 
(c) posterior density as a function of the LLLL and RRLL coupling cross sections.
\label{fig:fcnc_pol}}
\end{figure}

\subsection{Flavor-changing neutral currents}
The $t$-channel final state is sensitive to flavor-changing neutral currents via quark-gluon couplings.
In FCNC interactions a gluon vertex couples an up~quark or a charm~quark to the top quark. The
rate for these events is large because of the initial state containing two light quarks. The
\dzero~collaboration searched for FCNC interactions in the same dataset as used in the observation 
analysis~\cite{Abazov:2010qk}. 
Two FCNC couplings are considered, to up~quarks ($tgu$) and to charm quarks ($tgc$).
A Bayesian Neural Network (BNN) is used to separate the FCNC signal from the large backgrounds.
The BNN output distribution is shown in Fig.~\ref{fig:fcnc_pol}(a) and~~\ref{fig:fcnc_pol}(b).
The limits on the couplings are $\kappa_{tgu}/\Lambda < 0.013$~TeV$^{-1}$ and
$\kappa_{tgc}/\Lambda < 0.057$~TeV$^{-1}$, without making assumptions about the $tgc$ and $tgu$ couplings,
respectively. The corresponding limits on top quark decay branching fractions
are $\mathcal{B}(t \rightarrow gu)<2.0\times10^{-4}$ and $\mathcal{B}(t \rightarrow gc)<3.9\times10^{-3}$.

\subsection{Top quark polarization}
The observation of single top quark production assumes SM-like left-handed couplings of the top quark
to the $W$~boson and the $b$~quark. The CDF experiment performed a search for production of single
top quarks through right-handed couplings. The two specific models under investigation are SM-like
left-handed coupling at all four interaction vertices ($LLLL$, two for top quark production and two for
top quark decay) and right-handed anomalous couplings in top quark production together with 
left-handed couplings in top quark decay ($RRLL$). The Bayesian posterior probability density
as a function of the cross section for single top quark production with these two couplings is shown
in Fig.~\ref{fig:fcnc_pol}(c). The data are consistent with the SM, and the measured polarization 
is $\frac{\sigma_{RRLL} - \sigma_{LLLL}}{\sigma_{RRLL} + \sigma_{LLLL}} = -1^{+1.5}_{-0}$.

\section{Conclusions}
The Tevatron experiments \dzero~and CDF have each observed single top quark production at the
five standard deviation level and have measured the production cross section as well as
the CKM matrix element $V_{tb}$. The measurements as well as a measurement of the $t$-channel production mode
are consistent with the SM expectation.  
Several new physics scenarios have been explored in the single top quark final state and limits
have been set on a new $W'$~boson, on flavor-changing neutral currents, and on anomalous
top quark couplings.


\section*{References}
\begin{thebibliography}{99}
\bibitem{Abazov:2009ii}
  V.~M.~Abazov {\it et al.}  (D0 Collaboration),
  \Journal{\PRL}{103}{092001}{2009}.

\bibitem{Aaltonen:2009jj}
  T.~Aaltonen {\it et al.}  (CDF Collaboration),
  \Journal{\PRL}{103}{092002}{2009}.

\bibitem{Cabibbo:1963yz}
  N.~Cabibbo,
  Phys.\ Rev.\ Lett.\  {\bf 10}, 531 (1963).

\bibitem{Kobayashi:1973fv}
  M.~Kobayashi and T.~Maskawa,
  Prog.\ Theor.\ Phys.\  {\bf 49}, 652 (1973).

\bibitem{Group:2009qk}
  Tevatron Electroweak Working Group  (CDF Collaboration and D0 Collaboration),
  arXiv:0908.2171 [hep-ex] (2009).

\bibitem{Aaltonen:2010fs}
  T.~Aaltonen {\it et al.}  (CDF Collaboration),
  Phys.\ Rev.\  D {\bf 81}, 072003 (2010).

\bibitem{Abazov:2009nu}
  V.~M.~Abazov {\it et al.}  (D0 Collaboration),
  Phys.\ Lett.\  B {\bf 690}, 5 (2010).

\bibitem{stxs-kidonakis}
N.~Kidonakis,
Phys.\ Rev.\ D {\bf 74}, 114012 (2006).

\bibitem{stxs-harris} B.~W.~Harris {\it et al.}, Phys.\ Rev.\ D {\bf 66}, 054024 (2002);
Z.~Sullivan, Phys.\ Rev.\ D {\bf 70}, 114012 (2004);
J.~Campbell, K.~Ellis, and F.~Tramontano, {\it ibid.} {\bf 70}, 094012 (2004).

\bibitem{Tait:2000sh}
T.~Tait and C.-P.~Yuan,
Phys. Rev. D {\bf 63}, 014018 (2001).

\bibitem{Alwall:2007}
J.~Alwall {\it et al.}, Eur.~Phys.~J. {\bf C}~49, 791 (2007).

\bibitem{Abazov:2009pa}
  V.~M.~Abazov {\it et al.}  (D0 Collaboration),
  Phys.\ Lett.\  B {\bf 682}, 363 (2010).

\bibitem{Abazov:2010qk}
  V.~M.~Abazov {\it et al.}  (D0 Collaboration),
  Phys.\ Lett.\  B {\bf 693}, 81 (2010).

\bibitem{Abazov:2008yn}
  V.~M.~Abazov {\it et al.}  (D0 Collaboration),
  Phys.\ Rev.\ Lett.\  {\bf 100}, 192003 (2008).
 
\end{thebibliography}
\end{document}